# Investigation of RF performance of Ku-band GaN HEMT device and an in-depth analysis of short channel effects

Jagori Raychaudhuri[a,b], Jayjit Mukherjee[b,*], Sudhir Kumar[b], D.S. Rawal[b], Meena Mishra[b], Santanu Ghosh[a]

[a]*Department of Physics, Indian Institute of Technology, Delhi, Hauz Khas, Delhi-110016, India*

[b]*Solid State Physics Laboratory, DRDO, Timarpur, Delhi-110054, India*

*jyjt.dbb@gmail.com

**Abstract**

In this paper, we have characterized an AlGaN/GaN High Electron Mobility Transistor (HEMT) with a short gate length ($L_g \approx 0.15\mu m$). We have studied the effect of short gate length on the small signal parameters, linearity parameters and gm-gd ratio in GaN HEMT devices. To understand how scaling results in the variation of the above-mentioned parameters a comparative study with higher gate length devices on similar heterostructure is also presented here. We have scaled down the gate length but the barrier thickness($t_{bar}$) remained same which affects the aspect ratio ($L_g/t_{bar}$) of the device and its inseparable consequences are the prominent short channel effects (SCEs) barring the optimum output performance of the device. These interesting phenomena were studied in detail and explored over a temperature range of -40 °C to 80 °C. To the best of our knowledge this paper explores temperature dependence of SCEs of GaN HEMT for the first time. With an approach to reduce the impact of SCEs a simulation study in Silvaco TCAD was carried out and it is observed that a recessed gate structure on conventional heterostructure successfully reduces SCEs and improves RF performance of the device. This work gives an overall view of gate length scaling on conventional AlGaN/GaN HEMTs.

## 1. Introduction

GaN based technology is a boon in high power, high temperature and high frequency applications [1-3]. This wide band gap semiconductor promises variety of high efficiency and high voltage applications in the arenas of microelectronics, optoelectronics, telecommunications, satellite communications etc. [4-7]. Since next generation technology is in great demand of high frequency high speed applications, many studies are going on globally to achieve the desired high frequency performance in GaN based technology. Scaling down of gate length of the GaN HEMT device is the most popular way to achieve high frequency operation of the device [8-10]. Many studies have reported that GaN HEMTs with short gate length has shown promising DC and RF performances [11-13]. Although gate scaling caters impressive device performance in high frequency domain it also invokes an unavoidable phenomenon of the SCEs [14-17]. SCEs can significantly deteriorate the performance of GaN HEMTs as SCEs bar to achieve the expected frequency of operation, reduce gain, increase power dissipation and decrease device linearity. The efficiency and reliability of the device are also compromised. SCEs become observable when the aspect ratio (gate length (Lg)/ barrier thickness (tbar)) is not maintained properly

[16]. In the shorter Gate length devices as the channel length decreases, a number of detrimental effects arise. Drain Induced Barrier Lowering (DIBL), Channel Length Modulation (CLM), Gate Induced Drain Leakage (GIDL) are the prominent evidences of SCEs that hinder the output of the device. Drain induced barrier lowering (DIBL) is one of the important SCEs in short gate length devices operating in the subthreshold region. Due to this, the barrier between the source and the channel gets lowered with applied drain bias and injection of carriers into the channel increases subthreshold current which degrades the performance of the device [18-20]. In short channel devices, due to high electric field underneath the gate, profound band bending allows electron tunneling in the drain region and it increases the off-state leakage current known as gate induced drain leakage (GIDL) [21-23]. Channel length modulation (CLM) is another SCE which is clearly visible from the drain output characteristic curve of the short gate device. Due to the strong electric field the velocity saturation happens earlier at the drain side reducing the effective channel length, known as CLM. This leads to increase in drain current with increase in drain voltage even in the saturation region [17,24,25]. In previously reported literature we have seen how SCEs degrade the RF performance of HEMTs [15]. It is pointed out in literature [16] that aspect ratio has an in fluence on SCEs and degradation of RF performance. The SCEs were explored in Room Temperature [26] and the SCEs were modelled using 2D channel potential distribution [14]and an improved empirical Large Signal model was proposed including SCEs [17]. Various methods have been employed to suppress SCEs and to achieve better device performances. Barrier thickness can be reduced to improve the aspect ratio [16,27] but drastic reduction can affect the 2DEG density, hence InAlN barrier [28] became a choice over AlGaN for high frequency applications. DIBL can be reduced reportedly by using AlGaN back barrier [29]. Also Graded AlGaN buffer can be a solution for reduction of for SCEs [30]. But these demand growth challanges. Using negative Capacitance FINFET [31], using dual metal gate [32,33] the SCEs can be suppressed but these requires complex fabrication processes. Gate recess [34,35] can be another solution although an optimised approach should be used in this case to avoid the leakage

In this work we have characterized an in-house fabricated HEMT having a gate length of 0.15 µm over conventional AlGaN/GaN layer structure having an aspect ratio around 6. From different characterizations it is observed that the device yields good DC and RF performances. Small and large signal characteristics of the device denote that our device is well suited for high frequency applications. To support our findings comparison of performances were drawn with higher gate length devices on same heterostructure. On the other hand, with drastic scaling, the unavoidable signature of SCEs was also visible in our devices. These effects were investigated in detail over a temperature range. To suppress the effect of SCEs to a great extent without altering the conventional layer structure a simulation study was carried out where we have proposed a recessed gate structure to achieve our goal.

## 2. Sample and Experimental Details

The layer structure of the device and the top-view SEM image chosen for our experiment is shown in (Fig. 1). The device was fabricated on the conventional AlGaN/GaN heterostructure. The GaN buffer is grown on SiC substrate and the thickness of $Al_{0.25}Ga_{0.75}N$ barrier is around 25 nm. 1 nm of AlN spacer layer was used. The channel in the heterostructure has a charge sheet density of 1.1 x $10^{13}$ $cm^{-2}$ and electron mobility of 1930 $cm^2$/V.s. The device was passivated with a thin layer of PECVD grown SiN. 0.15 µm length and 75 µm gate width are chosen for fabricating T-gate and Ni/Au is used to achieve gate Schottky contact whereas Ti/Al/Ni/Au stacks were used to realize the drain and source ohmic contacts whose contact figure of merit (FOM) is around 0.6 Ω.mm.

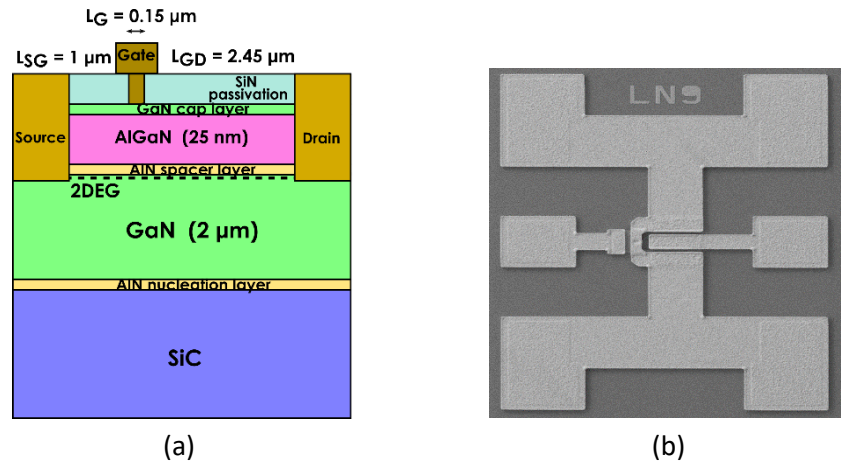

Fig. 1 (a) Layer structure of the device. (b) Top-view SEM image of the 0.15 µm gate length HEMT

DC-IV characterization was done using Keithley SCS 4200A semiconductor parametric analyzer. Output and transfer characteristics were obtained and the linearity parameters were calculated from the latter. Temperature dependent DC measurements were carried out in the range of -40°C to 80°C. The signature of SCEs was also evaluated in detail. S-parameter measurements were also carried out at different bias conditions to estimate RF performance of the device. From the measured S-parameter data $f_T$ and $f_{max}$ were calculated and intrinsic device parameters were extracted following small signal equivalent circuit model. The SCEs and RF performance of the short gate length device was compared with higher gate length devices fabricated on same layer structure. The power measurement was conducted by Load Pull system terminated at maxPout. The system consists of a signal source, pre amplifier, isolator, bias tee, directional coupler and tuners to measure the power levels. The measurements were done at CW mode at 16 GHz. Simulation study was conducted in Silvaco TCAD to realize the recess structure to control SCE after calibrating DC-IV and transfer characteristics on the 0.15 µm gate length device.

### 3. Experiment Results:

### 3.1 DC and RF Characterization of the Device

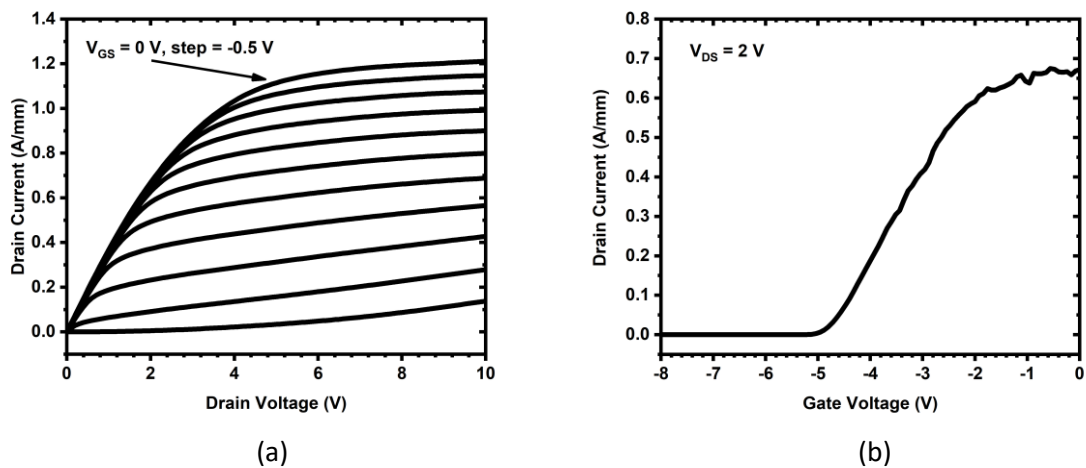

Fig. 2 DC-IV characteristics of the 0.15 µm gate length device (a) Output characteristics (b) Transfer characteristics measured at VDS = 2 V.

From the DC-IV output characteristics, it is found that the device exhibits high saturation current (≈ 1.2 A/mm) (Fig. 2). From the transfer characteristic the maximum transconductance, $gm_{max}$ is obtained as 255 mS/mm. The transconductance (DC gm) can be expanded into a power series [36] with respect to VGS. The power series consists of gm, gm', gm" where, gm' is the first order derivative of the transconductance (gm) with respect to the gate voltage (VGS) whereas gm'' is the second order derivative of the transconductance (gm) with respect to the gate voltage (VGS). Device linearity depends on the flatness of the derivatives of gm with VGS. The lower magnitudes of gm', gm" show good linearity [37]. Fig3 shows a comparison of gm, gm', gm" of three devices having different gate lengths grown on same heterostructure and all of three have gate width of 75 μm. The reduced values of gm', gm'' estimate good linearity in 0.15 μm device.

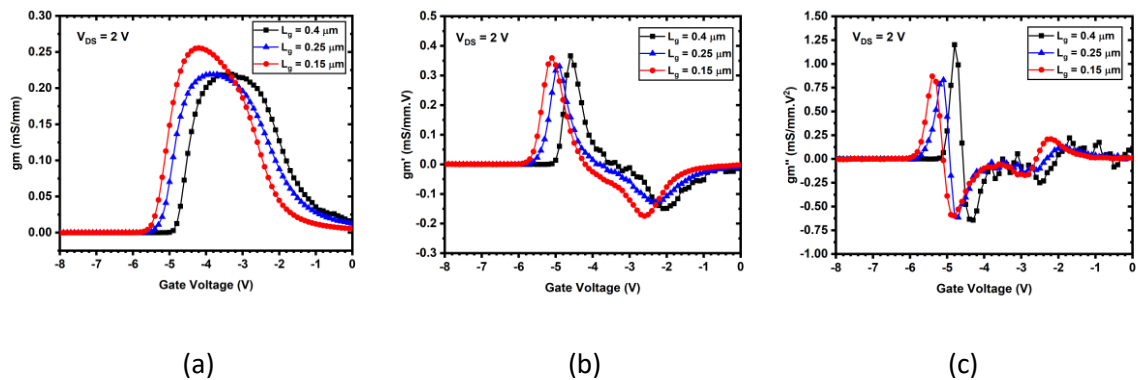

(a) (b) (c)

Fig. 3 (a) gm (b) gm' and (c) gm'' variation with gate voltage for three different gate length GaN HEMTs.

Now, the device parameters of our short gate length device were extracted from small signal equivalent circuit using measured S-parameters [38,39,40]. After calculating the extrinsic parameters, the intrinsic parameters were obtained. To extract extrinsic capacitances such as gate pad capacitance and drain pad capacitance (Cpg and Cpd respectively) the S-parameters were measured at VGS = -4.6 V and VDS = 10 V, to extract extrinsic inductances and resistances the S- parameters were measured at VGS = 0 V and VDS = 0 V, whereas to extract the intrinsic parameters the S-parameters were measured at VGS = -3 V and VDS = 10 V. The comparison of the device parameters with those of devices with larger gate length is enlisted in Table 1.

| | Parameters | 0.4 μm | 0.25 μm | 0.15 μm |
|---|---|---|---|---|
| **Extrinsic** | $R_g$ (Ω) | 1.51 | 3.7 | 2.08 |
| | $R_d$ (Ω) | 8.11 | 5.5 | 7.75 |
| | $R_s$ (Ω) | 6.31 | 4.79 | 3.55 |
| | $L_d$ (pH) | 97.5 | 98 | 96.72 |
| | $L_g$ (pH) | 73 | 78 | 109.3 |
| | $L_s$ (pH) | 3.9 | 3.34 | 6.68 |
| | $C_{pg}$ (fF) | 15.95 | 15.9 | 16.2 |
| | $C_{pd}$ (fF) | 28.8 | 26.88 | 36.8 |

|  | $g_m$ (mS) | 52.75 | 59 | 67 |
|---|---|---|---|---|
| **Intrinsic** | $g_d$ (mS) | 3.7 | 5 | 6.72 |
|  | $R_i$ (Ω) | 3.98 | 3.1 | 2.7 |
|  | $C_{gd}$ (fF) | 29.6 | 28.8 | 22.8 |
|  | $C_{gs}$ (fF) | 238 | 236 | 210 |
|  | $C_{ds}$ (fF) | 75.26 | 75.4 | 47.4 |
|  | τ (ps) | 1.42 | 1.32 | 1.26 |

Table 1 Comparative device parameters using small signal equivalent circuit model for three different gate length devices.

It can be seen that the 0.15 μm gate length device shows higher value of RF transconductance (RF gm) indicating better current gain, lower value of transit time (τ) represents faster charging time and decreased value of intrinsic capacitances denotes its suitability in high frequency applications.

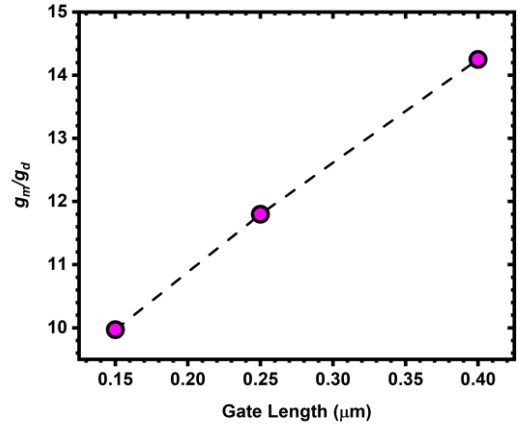

Fig. 4 $g_m$(RF gm)-$g_d$ ratio variation with gate length for GaN HEMTs

Fig. 4 shows lower value of $g_m$-$g_d$ ratio in the case of the 0.15μm gate length device compared to the other devices. In spite of having better RF gain, this ratio degrades due to higher value of output conductance ($g_d$). It is always desirable to get lower value of $g_d$ to achieve better RF performance of the device but here the high value of drain conductance in 0.15μm gate length device indicates the influence of short channel effects which has been discussed in Section II.

The RF performance of the device was also examined by calculating cut-off frequency ($f_T$) and maximum oscillation frequency ($f_{max}$) from unity current gain ($h_{21}$) and Maximum gain (MAG/MSG) using S-parameter measurements at the bias point ($V_{GS}$ = -3 V, $V_{DS}$ = 10 V). It is obtained that $f_T$ = 36.1 GHz and $f_{max}$ = 40.5 GHz for the 0.15 μm device from Fig. 5.

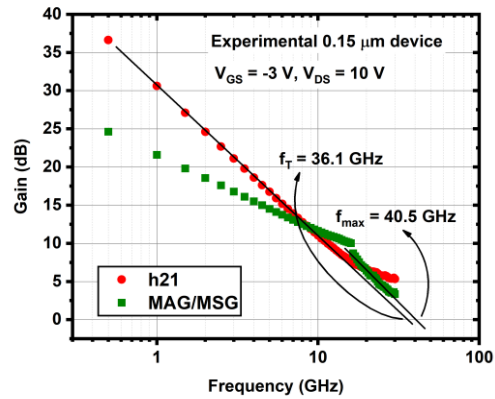

Fig. 5 f$_T$ and f$_{max}$ as calculated from the unity current gain (h$_{21}$) and maximum gain (MAG/MSG) of the 0.15 µm HEMT.

| Gate Length (µm) | f$_T$ (GHz) | f$_{max}$ (GHz) |
|---|---|---|
| 0.4 | 26 | 32 |
| 0.25 | 31 | 35 |
| 0.15 | 36.1 | 40.5 |

Table 2 Comparative f$_T$ and f$_{max}$ for three different gate length devices.

As a comparison f$_T$ and f$_{max}$ were also calculated for a 0.4 µm and a 0.25 µm gate length device fabricated on same heterostructure. It is seen from Table 2 that the 0.15 µm device shows higher frequency of operation.

Large signal power performance was carried out by using load pull technique in CW mode at V$_{DS}$ = 24 V, V$_{GS}$ = -4.2 V. From this we have obtained a saturated power density of 2.6 W/mm at 16 GHz showing its suitability in the Ku-band depicted in Fig. 6.

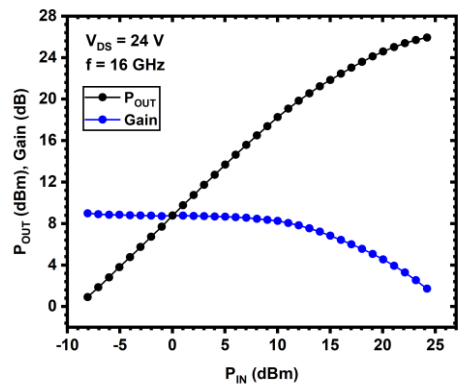

Fig. 6 Large signal power performance of the 0.15µm gate length HEMT

## 3.2 Investigation of SCEs

It is reported in literature that 0.15 µm gate length devices yield $f_T$ of 50 GHz [12,13], 79 GHz [41], 85 GHz [42] and gm of 281 mS [12,13], 288 mS [43], 400 mS [44], 260 mS [26]. Comparing our device performance with the reported data we have observed that our device has further scope of improvement although there are some variations in layer structure, material and processing with respect to the reported ones. SCEs are the prominent reasons that can be responsible for not getting the expected output.

When we scale down the gate length of the device, we observe the SCEs like DIBL, GIDL, CLM etc. which hinder the optimum performance of the device. From the output DC-IV characteristics of our device, it is observed that output current does not saturate at higher $V_{DS}$ values. This gives us the evidence of CLM in our device already observed from Fig. 2. This phenomenon is also supported from $g_d$ calculated in Table 2. To further explore other SCEs, DIBL of the device is calculated for $V_{DS}$ = 2 V and 1 V at the subthreshold region. The DIBL for the 0.15 µm device is calculated as 140 mV/V. Now the same was calculated for the 0.4 µm and 0.25 µm gate length devices.

| Gate Length (µm) | DIBL (mV/V) |
|---|---|
| 0.4 | 50 |
| 0.25 | 70 |
| 0.15 | 140 |

Table 3 Comparative study of DIBL for the three different gate length devices.

Table 3 shows the comparative values of DIBL for these three devices from which we can say the effect of DIBL is pretty high in the short gate length device as a side effect of scaling. To explore this effect in detail we have conducted temperature dependent study of the SCEs. Subthreshold swing (SS) and DIBL were measured over a range of -40$^O$C and 80$^O$C and their variation is illustrated in Fig. 5.

The SS shows a monotonic increasing trend with temperature (Fig. 7(a)). The DIBL however shows an interesting trend where it increases above 0$^O$C as well as below 0$^O$C. The increase above 0$^O$C is attributed to thermionic emission [45]. In the cryogenic regime, a similar phenomenon was observed in the case of short channel NMOS device where the DIBL increased below 0$^O$C. It was reported that the effective channel length becomes shorter when the device operates in the cryogenic temperature [46]. The shortening of the channel length can be extracted from source-drain potential profile. This phenomenon becomes more profound in the devices already having shorter gate lengths. Hence for our device, we can observe increased DIBL with decreasing temperature. This effect is also supported from higher offstate leakage values in this temperature range (inset of Fig. 7(b)). However, in our device the evidence of GIDL is not so prominent.

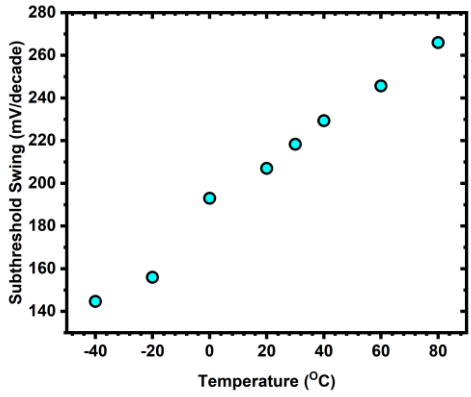 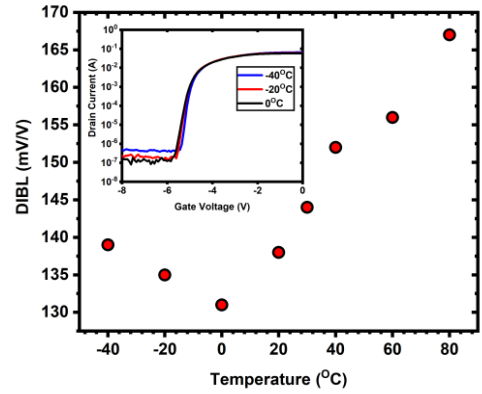

(a)                      (b)

Fig. 7 Subthreshold swing (SS) with temperature and DIBL variation with temperature for the 0.15µm gate length device. The inset of (b) shows the off-state drain leakage in the cryogenic regime.

### 3.3 Simulation study to suppress SCEs

So far, we have observed that the 0.15µm gate length GaN HEMT device fabricated on conventional layer structure is well suited for high frequency applications but due to the aggressive scaling it suffers from SCEs that obstructs it to achieve optimum performance. So, we have aimed to suppress the SCEs to great extent.

We have carried out a simulation study in Silvaco TCAD to reduce the SCEs without altering the heterostructure and employing a recessed gate. To begin with we have calibrated the experimental DC-IV (both transfer and output) characteristics of our device. The layer structure and device dimensions were kept same as our physical device shown in Fig. 1. The physical models that were used are Fermi statistics, Shockley-Read-Hall (SRH) recombination and GaNSat for mobility. The fine tuning of the 2DEG density and electron mobility was performed to achieve proper calibration. Fig. 8 shows the calibrated characteristics.

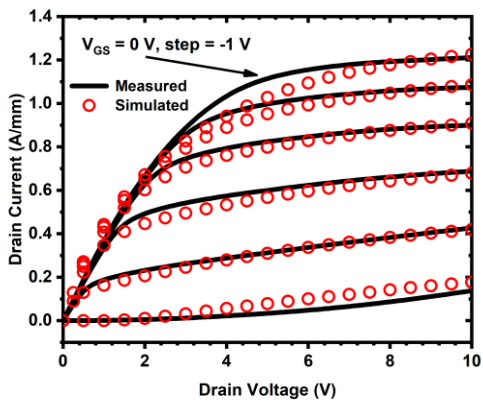 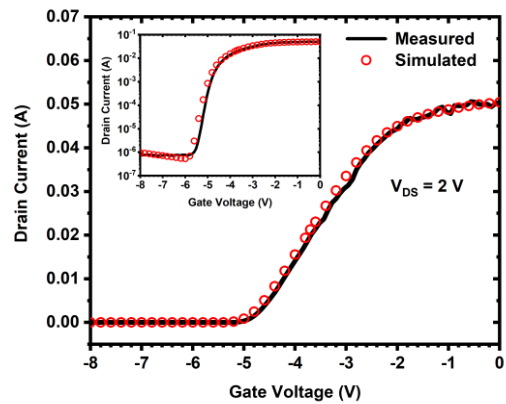

(a)                      (b)

Fig. 8 Calibration of experimental device characteristics with simulated results for the 0.15 µm HEMT

Keeping the layer structure and physical models same we have employed the recessed gate structure into the AlGaN barrier. For optimization, a variation of DIBL with recess depth is depicted in Fig. 9(a). It is reported that with recess depth the SCEs are neglected but on the flip side it also leads to increased gate leakage [31]. So, we have optimized a recess depth of 10 nm (Fig. 9(b)) by incorporating which we have reduced DIBL at 30 mV/V (measured for VDS = 1 V and VDS = 2 V) i.e. 78.57% reduction with respect to the experimental DIBL obtained without aggravating the gate leakage ($10^{-6}$ A).

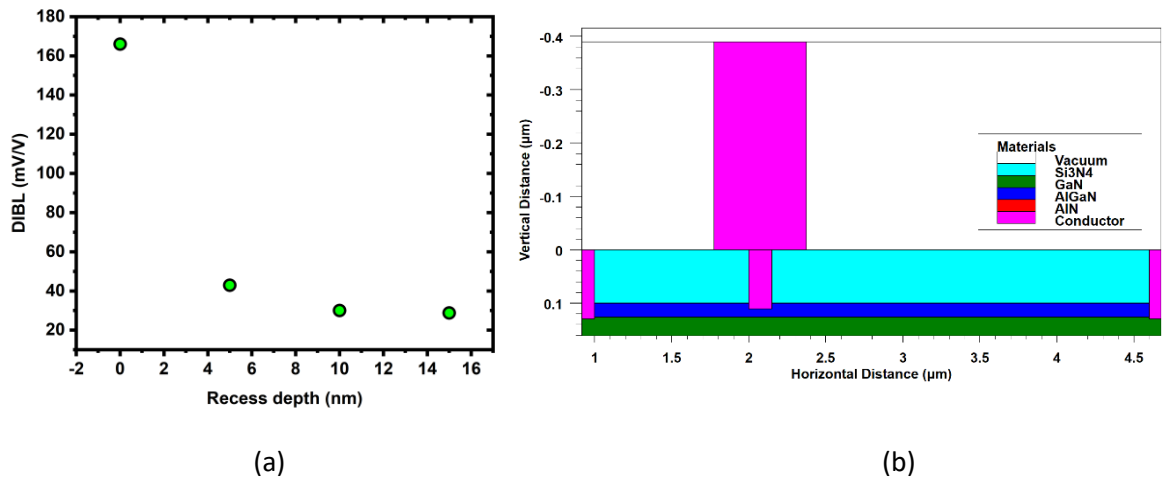

(a) (b)

Fig. 9 (a) DIBL trend with recess depth for 0.15 μm gate length device. (b) Schematic of the optimized recess depth of 10 nm T-gate device.

The DIBL simulated is at per the experimental DIBL of the 0.4 μm gate length device. DIBL values of GaN HEMT are reported as 45mV/V for 90 nm gate length [47], 26mV/V 100 nm gate length [48], 12 mV/V for 140nm gate length [49], 100 nm 40mV/V for 100 gate length [50]. Observing these results, it can be concluded that our device shows improvement after recess.

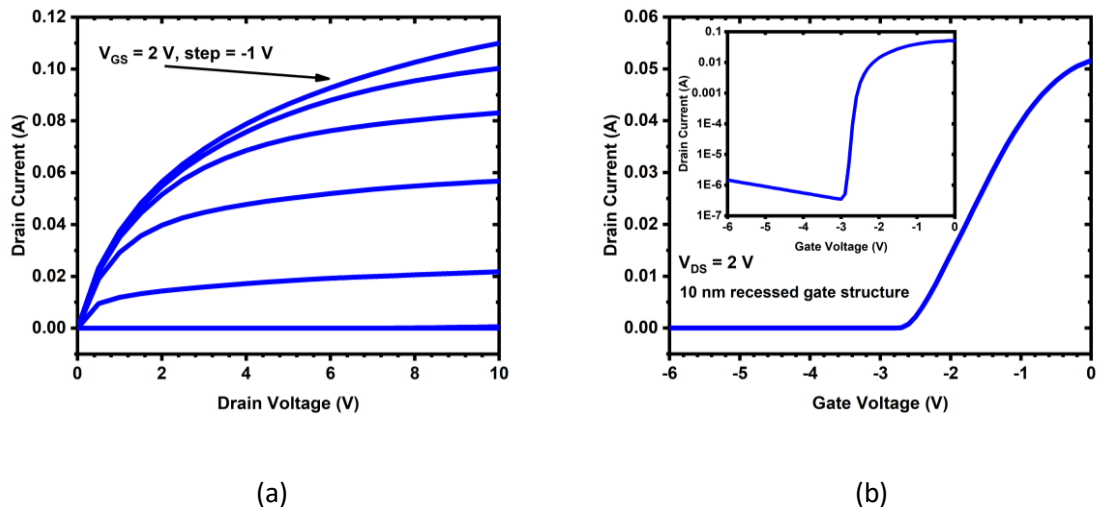

(a) (b)

Fig. 10 Simulated DC-IV characteristics of the 10 nm recessed 0.15μm gate length HEMT. (a) Output characteristics. (b) Transfer characteristics with inset shown in logarithmic scale.

From the drain output characteristics, it can be seen that the CLM has also been reduced and the drain current has also increased Fig. 10(a). Also, we have observed from Fig. 10(b) that the threshold voltage $V_T$ has increased to -2.5 V that is the device is going towards the enhancement mode operation. This leads to delayed saturation at higher drain biases for VGS > 0 V. But in this work our investigation was focused up to VGS = 0 V and there was no signature of CLM in the recessed gate HEMT upto VGS = 0 V. Along with that the RF performance of the 0.15µm gate length recess structure device has improved. It is reflected as we have obtained $f_T$ = 50.4 GHz and $f_{max}$ = 88.8 GHz in our simulated recessed structure (Fig. 11).

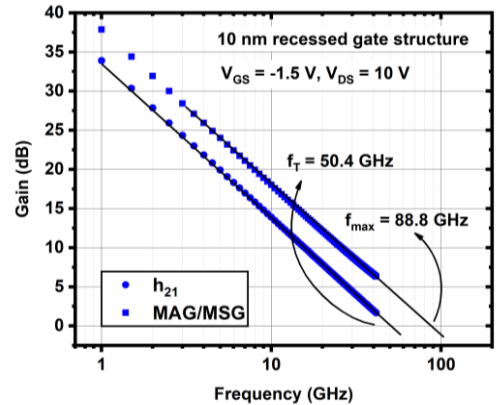

Fig. 11 Simulated $f_T$ and $f_{max}$ as calculated from the unity current gain ($h_{21}$) and maximum gain (MAG/MSG) of the 10 nm recessed 0.15 µm HEMT.

A comparative study of RF performance of GaN HEMTs having 0.15 µm gate length is enlisted in Table 4. From this we can see that the 0.15 µm gate length device is showing good RF performance although not the best since it is affected by SCEs but it improves after curbing these detrimental effects by gate recess.

| $f_T$ (GHz) | $f_{max}$ (GHz) | Reference |
|---|---|---|
| 50 | 121 | [13] |
| 41 | 70 | [51] |
| 47.3 | 97 | [52] |
| 74.2 | - | [53] |
| 29.8 | 50.9 | [54] |
| 36.1 | 40.5 | This work |
| 50.4 | 88.8 | Recessed gate simulation (this work) |

Table 4 Comparative $f_T$ and $f_{max}$ for different reported literature on 0.15 µm GaN devices.

## 4. Conclusion

In this article, we have demonstrated a 0.15µm gate length device fabricated on conventional layer structure. In spite of having a low aspect ratio (≈ 6) for GaN HEMT our device has shown good DC and RF performances. Intrinsic device parameters like $g_m$, τ and intrinsic capacitances ($C_{gs}$, $C_{gd}$) have indicated towards better RF operation. Without altering the basic layer structure, we have achieved appreciable amount of cut-off frequency, gain and power density at 16 GHz making it eligible for Ku-band operation. SCEs like CLM and DIBL were observed and explored in our device, although the presence of GIDL was negligible. The behaviour of DIBL and Subthreshold slope with temperature has been studied. The DIBL in cryogenic regime increased which was attributed to the shortening of effective channel length. Exploring SCEs over a temperature range can give an insight in further modelling of GaN HEMTs. From simulation a 10 nm recessed T-gate structure was proposed on the conventional layer structure to reduce the effect of DIBL and to improve $f_T$ and $f_{max}$ of the device. In the recessed structure we have observed a DIBL comparable to that of fabricated 0.4 μm device. This gives further scope to study recessed gate structure on conventional GaN epilayers.

**CRediT authorship contribution statement**

**Jagori Raychaudhuri:** Conceptualization, Formal Analysis, Methodology, Software, Writing -original draft. **Jayjit Mukherjee:** Conceptualization, Formal Analysis, Data curation, Software. **Sudhir Kumar:** Data Curation. **D S Rawal:** Conceptualization, Resources, Writing – review & editing. **Meena Mishra:** Supervision, Validation, Writing – review & editing, Project administration. **Santanu Ghosh:** Supervision, Validation, Writing – review & editing.

**Declaration of competing interest**

The authors declare that they have no known competing financial interests or personal relationships that could have appeared to influence the work reported in this paper.

**Acknowledgment**

The authors would like to thank GaN Team and Director SSPL, DRDO, Delhi for the valuable support and guidance to carry out the experimental work.